\documentclass[%
 reprint,
superscriptaddress,
prl,
nofootinbib,
 amsmath,amssymb,
 aps,
]{revtex4-2}
\usepackage{xcolor}
\usepackage{graphicx}
\usepackage{dcolumn}
\usepackage{bm}

\newcommand{\mnras}{MNRAS}

\newcommand{\pasa}{PASA}
\newcommand{\araa}{ARA\&A}
\newcommand{\apjl}{ApJL}

\begin{document}

\preprint{APS/123-QED}

\title{Plasticity of Neutron Star Crusts}

\author{Matthew E. Caplan}
\email{Corresponding author: mecapl1@ilstu.edu}
\affiliation{Department of Physics, Illinois State University, Normal, IL 61790, USA}
\affiliation{Department of Physics, University of Illinois Urbana–Champaign, Urbana, IL 61801, US}

\author{Nevin T. Smith}
\affiliation{Department of Physics, Illinois State University, Normal, IL 61790, USA}

\author{Ashley J. Bransgrove}
\affiliation{Princeton Center for Theoretical Science and Department of Astrophysical Sciences, Princeton University, Princeton, NJ 08544, USA}

\author{Charles J. Horowitz}
\affiliation{Center for the Exploration of Energy and Matter and Department of Physics,
Indiana University, Bloomington, IN 47405, USA}


\date{\today}

\begin{abstract}

We use first-principles molecular dynamics simulations to study the deformation and breaking of neutron star crusts. When simulating with strain rates several orders of magnitude slower than prior work, we find a new regime of steady plastic flow beyond the breaking point that is independent of the initial crystal structure. Polycrystals exhibit a robust transition from linear elasticity to perfect plastic flow at shear strains of $\epsilon = 0.05$, while monocrystals break at $\epsilon = 0.11$ and then flow plastically. The universal post-break plasticity may arise because the crystal self-consistently assumes a defect density to accommodate the imposed strain rate. If broken crusts can re-anneal to large crystal sizes, crust breaking may repeat with implications for magnetar bursts and flares.

\end{abstract}

\maketitle


\textit{Introduction\textemdash} Neutron stars rapidly cool after formation until the plasma in their outer layers freezes into a crystalline solid \cite{Ruderman1968}. The elastic properties of this crystal, which spans densities up to $10^{14} \, \rm g \, cm^{-3}$, regulate a host of observable astrophysics including starquakes and magnetar outbursts \cite{Blaes_1989B, Thompson_1995,Thompson_1996, Beloborodov_2014, Bransgrove_2020}. Mountains supported by elastic stresses on rotating neutron stars may also be an observable source of continuous gravitational waves \cite{Lasky2015GWNS,Riles2023continuous,Haskell2023continuous}. Some elastic properties of this crystal are known, especially in the linear regime that is accessible with theory and small simulations \cite{Ruderman1968,horowitz2009breaking,Chugunov2010,Hoffman2012,Baiko2011,KobyakovPethick2014,Kobyakov2015,Chugunov2021,Kozhberov2022,Kozhberov2023,Zemlyakov2025}, but the behavior of larger crystals at very large strains is less well understood despite their importance to astrophysics.

Molecular dynamics (MD) simulations enable first principles modeling of deformed neutron star crusts to extract transport properties, and have successfully verified the shear modulus and some information about the breaking strain ($\epsilon \approx 0.11$) for perfect crystals \cite{horowitz2009breaking,Chugunov2010,Hoffman2012}. The post-break dynamics are less well understood as these are strongly dependent on the exact crystal defect network and require large simulations run for long times. While polycrystal elasticity has been studied with analytics \citep[e.g.][]{Kobyakov2015,BaikoChugunov2018} it is almost entirely unstudied with MD with a notable exception of one simulation in \citet[hereafter HK09]{horowitz2009breaking}.
If the initial crust forms as a polycrystal of many grains, then the study of polycrystals is immediately well motivated. 
However, even if the crust initially consists of macrocrystals \cite{Baiko2024}, one might reasonably argue that a transient involving crust breaking will introduce many crystal defects that freeze into a complicated network of grains.

In this work, we verify that the neutron star crust is elastic up to the breaking strain, but by simulating with strain rates several orders of magnitude slower than in prior work we find a converged regime of perfect plasticity out to very large strains.

\textit{Molecular Dynamics\textemdash} 
The neutron star crust consists of a fully pressure ionized plasma of nuclei with pressure support from electron degeneracy pressure (and neutron degeneracy pressure at densities above neutron drip). 
At sufficiently low temperature, nuclei arrange into a body-centered cubic (bcc) lattice that freezes at $\Gamma \gtrsim 175$ where $\Gamma = e^{2} Z^{2} / a_i T$ where $T$ is temperature, $eZ$ is the nuclear charge, and $a_i = (4\pi n_i / 3)^{-1/3}$ is the ion Wigner-Seitz radius \cite{baiko2022ab} with ion number density $n_i$.

Nuclei are therefore treated classically using the two-body repulsive Coulomb potential with Yukawa screening:
\begin{equation}
    V(r_{ij}) = \frac{e^2 Z_i Z_j}{r_{ij}} \exp\left(-\frac{r_{ij}}{\lambda}\right),
\end{equation}
where $r_{ij}$ is the separation between nuclei with charges $eZ_i $ and $eZ_j$, and $\lambda$ is the screening length from the relativistic Thomas–Fermi approximation ${\lambda^{-1} = 2k_F \sqrt{\alpha / \pi}}$. The dimensionless screening parameter is $\kappa = a_i / \lambda$. Using electron Fermi momentum $k_F = (3\pi^2 n_e)^{1/3}$ and electron density $n_e = \langle Z \rangle n_i$, this approximation is valid for densities above about $10^8 \ \rm{g/cm^3}$. Simulations do not include electrons explicitly; as a uniform relativistic gas they participate only through the screening, which simplifies to $\kappa(Z) = 0.185 \langle Z \rangle^{1/3}$. Throughout this work we use $Z= 40$ and $A=150$ ${(\kappa(Z) = 0.63)}$.  
We use periodic boundaries and cubic volumes and a timestep of $\tau = \omega_p^{-1}/17$ (ion plasma frequency ${\omega_p = \sqrt{4\pi e^{2} Z^{2} n_i / m}}$).

A volume-preserving shear deformation is applied to the periodic box boundaries, transforming the cube into a parallelepiped. The time dependent shear strain $\epsilon_{xz}(t)$ is applied in the $xz$-plane with a constant strain rate such that $\epsilon_{xz}(t) = \dot{\epsilon} \, t$. 
This differs slightly from the `frozen layers' to induce deformation used in HK09 that fixes and translates the initial crystal configuration in thin layers at the boundaries. Our method has the advantage of not imposing perfect crystal structure or seeding specific grain boundaries at the box boundary and so the response is self-consistent without bias from the initial conditions. Stress tensor elements including $\sigma_{xz}$ are calculated from pairwise forces and separations.  

These simulations are computationally challenging due to the infinite range of the Coulomb force, the large particle numbers needed to simulate a sufficiently large number of grains to have a polycrystals, and the long simulation times needed for convergence. Advances in GPU supercomputing and algorithms, especially in the classical MD code LAMMPS, enable our new generation of calculations \cite{thompson2022lammps}. Strain rates $(\dot\epsilon \approx 10^{-5} \omega_p)$ and cut-off distances for the Coulomb potential ${(r_{\rm cut} \approx 10 \lambda)}$ necessary for convergence are now known in the literature, making our approach tractable \cite[\textit{e.g.}][]{Chugunov2010,Hoffman2012}.

This deformation is only quasi-static if the ratio of the box deformation speed ${v_{\mathrm{box}} = \dot{\epsilon} L}$ to the elastic wave speed  $v_{\rm el} = \sqrt{\mu/\rho}$ is small, such that there is time for elastic information to be communicated across the box. Using $L = (N/n_i)^{1/3}$ the box size for $N$ nuclei, $\rho = n_i A$, and elastic modulus $\mu \simeq 0.11\, n_i (Ze)^2/a_i$, one obtains
\begin{equation}
\frac{v_{\mathrm{box}}}{v_{\rm el}} \simeq 8.5\, N^{1/3}\, \frac{\dot{\epsilon}}{\omega_p}.
\end{equation}
Observe that even for a seemingly slow $\dot\epsilon$, larger simulation volumes approach hydrodynamic shock. The polycrystal in HK09 contains $N=12.8\times10^6$ with an implied $\dot\epsilon = 1.08 \times 10^{-4} \ \omega_p^{-1}$, such that ${v_{\mathrm{box}}}/{v_{\rm el}}=0.21$ near the shocked regime. This likely explains the strain softening and ``amorphization'' reported in that work \cite{Caplan2026}.%

\begin{figure}[]
\includegraphics[trim={7 9 18 16},clip,width=0.98\columnwidth]{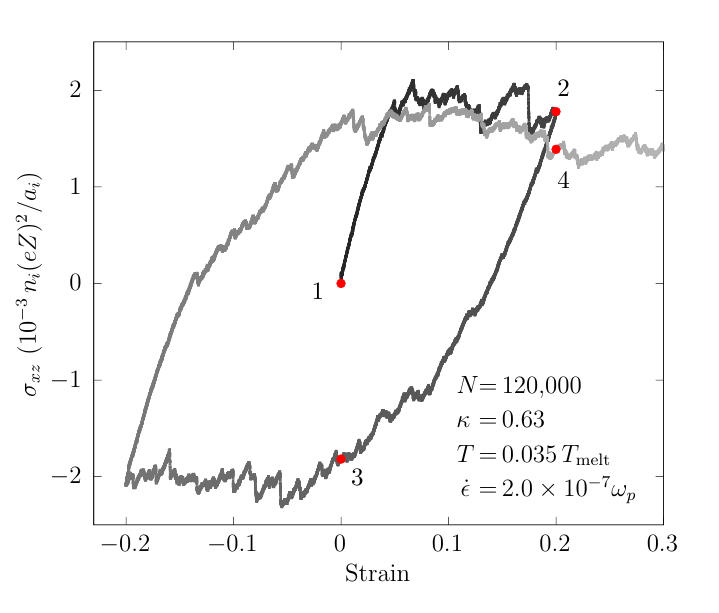}
\includegraphics[width=0.98\columnwidth]{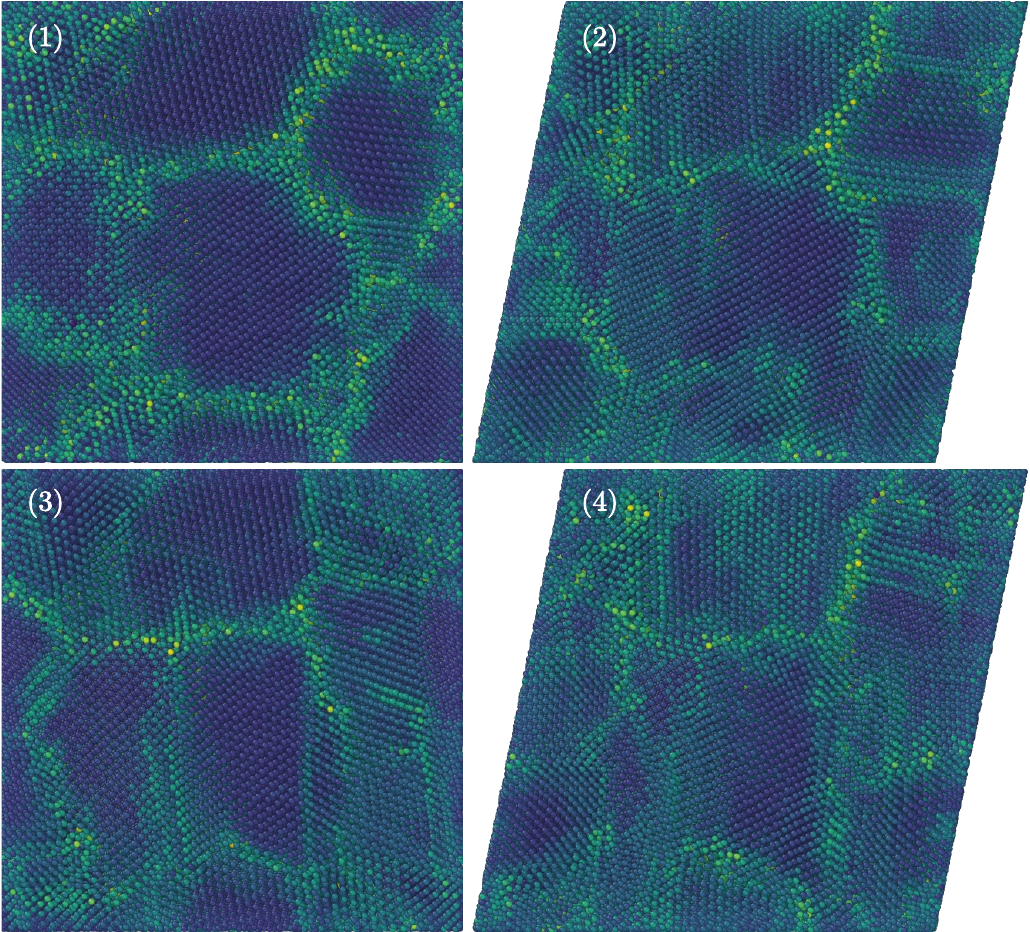}
\caption{\label{fig:hysteresis} Elastic hysteresis curve demonstrates perfect plasticity (top, color shows time), with configurations (bottom) showing the irreversible evolution of the grains and defects. Nuclei are colored by their bond order parameter $Q_6$ with lighter (darker) colors indicating particle disorder (order) with respect to nearest neighbors.}
\end{figure}

We use an NVT thermostat in all runs; in real neutron stars any heating from deformation should be quickly conducted away by electrons \cite{Tuckerman2006NVT}. To equilibrate before boxes are deformed our simulations are evolved at constant temperature until the energy is constant within noise.

\textit{Simulations\textemdash } 
We begin by exploring loading and unloading behavior of a polycrystal with an elastic hysteresis curve in Fig. \ref{fig:hysteresis}. 
The initial polycrystal contains $N=120\,000$ nuclei in 12 grains of roughly equal size. The typical grain diameter is about 17 lattice spacings; simulations of perfect crystals demonstrate convergence at this scale \cite[e.g. Fig 1 in][]{Hoffman2012}. This is sufficiently large so that grain boundaries on opposite sides of a grain are beyond the potential cut-off radius while maximizing the volume filling fraction of grain boundaries in our simulation.

We simulate at $\Gamma = 5000 \ (T\approx 0.035\,T_{\rm melt})$ and strain to ${\epsilon = 0.2}$ over $\tau=10^6 \ \omega_p^{-1}$ for ${\dot{\epsilon} = 2\times10^{-7} \omega_p}$ ${(v_{\rm box}/v_{\rm el} = 8.5\times10^{-5})}$. 
The stress increases almost linearly with strain up to $\epsilon = 0.05$, consistent with past work. We find a yield stress of about ${\sigma_{\rm yield} = 2\times10^{-3} \, n_i (eZ)^2/a_i}$. Beyond $\epsilon>0.05$ the stress remains constant, consistent with elastic-perfectly plastic materials \cite{negahban2012mechanical}. This is qualitatively new behavior that was not resolved in HK09. The variations of $\Delta \sigma \approx 0.1 \sigma_{\rm yield}$ in stress about the yield stress is due to stochastic snaps and localized yielding events. In a much larger box with many more grains the relative contribution of snaps to $\Delta \sigma$ will be much smaller so $\sigma$ will be constant. Preliminary analysis of this crackling finds that the snap magnitudes follow a power law; this suggests scale-invariant avalanche behavior and will be presented in future work \cite{sethna2001}.

An animation of this simulation is given in the supplemental materials (SM1).\footnote{Preprint only: Animations available in the arXiv ancilliary files.} It is challenging to track the evolution of the microscopic networks of defects, as there are no true bonds like in terrestrial materials. Instead, we visualize the crystal using the local bond order parameter $Q_6$ ($Q_6\approx 0.5 \,  (0.2)$ for a solid (liquid), giving high contrast between bulk crystal and defects \cite{steinhardt1983bond,Caplan2020structure}). This gives similar information to the von Mises strain. 
In the linear elastic regime we observe that the deformation of the box does not appear to have much effect on the crystal structure, but small deviations from linear elasticity from microslips of defects and grain boundaries could be a creep mechanism, warranting detailed future study. Once in the plastic regime defects proliferate. 
When unloading from $\epsilon=0.20$, the stress returns to zero nearly linearly by $\epsilon \approx 0.15$, indicating a permanent plastic strain of 15\%. This demonstrates that the crust may undergo irreversible deformation beyond the canonical breaking point without hardening. We interpret this microscopically: during the initial phases of loading, the linear elastic deformation strains the bcc unit cell of the bulk crystal to about $\epsilon=0.05$, beyond which the stress-activated plastic flow sets in facilitated by defects, especially those already present in grain boundaries. The bcc unit cells in the bulk remain at this constant linear deformation for the entire duration of plastic flow. When unloaded, the observed linear elastic deformation is the relaxation of this strain in the bulk bcc crystal. During unloading, the linear trend continues to about box strains $\epsilon \approx 0.10$, roughly consistent with our interpretation of the bulk crystal reaching a maximum elastic strain of $0.05$, but now in the opposite direction. For $0.10\gtrsim \epsilon \gtrsim -0.05$ we observe some modest strain hardening. This may be due the defects reorienting to support plastic flow in the opposite direction. When the loop is completed from $\epsilon=-0.20$ to $\epsilon = 0.20$ we observe similar hardening. The maximum stress in the plastic regime may be reduced by about ten percent during the final loading. The number density of defects may have grown, reducing the strength of the sample. 

In the final configuration (Fig. \ref{fig:hysteresis}d), the distinction between the grain boundaries and the bulk is less well defined and defects are distributed throughout the bulk. It could be that grain boundaries radiate dislocations into the bulk, or that many new defects are also nucleating to provide the necessary slip to accommodate the imposed strain rate. A more detailed analysis of dislocations will be presented alongside the crackling analysis in future work.

In the plastic regime, the crystal mobility should be governed by dislocations \cite{Orowan_1940}. The Orowan equation gives ${\dot{\epsilon} \propto b \, n \, v}$, with $b$ the magnitude Burgers vector, and $n$ and $v$ the number density and velocity of dislocations. If the Burgers vector can be treated as a grain averaged constant, then the product $nv$ is constant for a given $\dot\epsilon$. At finite temperature, dislocation motion has activation physics with a velocity of the form ${v(\sigma) = v_{0} \exp(- \Delta G(\sigma)/kT )}$, where $\Delta G(\sigma)$ is an activation barrier that decreases with stress. This barrier is commonly $\Delta G(\sigma) = \Delta G_0 (1 - |\sigma|/\sigma_c)^p$, with $\Delta G_0$ the zero-stress activation energy, $\sigma_c$ a critical stress, and $p$ a constant. If the barrier vanishes at $\sigma_c = \sigma_{\rm yield}$, then $\Delta G \to 0$ and dislocation motion becomes effectively athermal with viscous drag that scales with stress.  If sensitivity to temperature is subdominant, then all that is required for perfect plasticity is a steady state dislocation density that accommodates the imposed strain rate in the plastic regime. The elastic-perfectly plastic behavior therefore implies an asymptotic $\sigma_{\rm yield}$ at low temperature and small $\dot\epsilon$, motivating our next simulations.

\begin{figure*}
\includegraphics[trim={11 14 21 16},clip,width=0.33\textwidth]{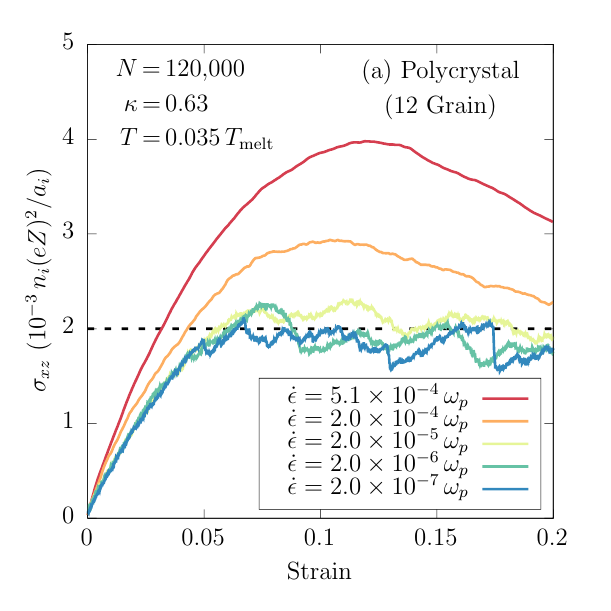}%
\includegraphics[trim={11 12 18 17},clip,width=0.33\textwidth]{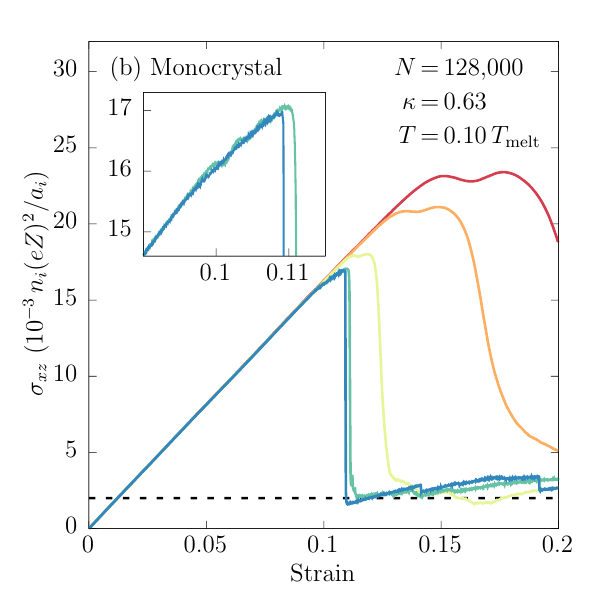}%
\includegraphics[trim={11 12 18 16},clip,width=0.33\textwidth]{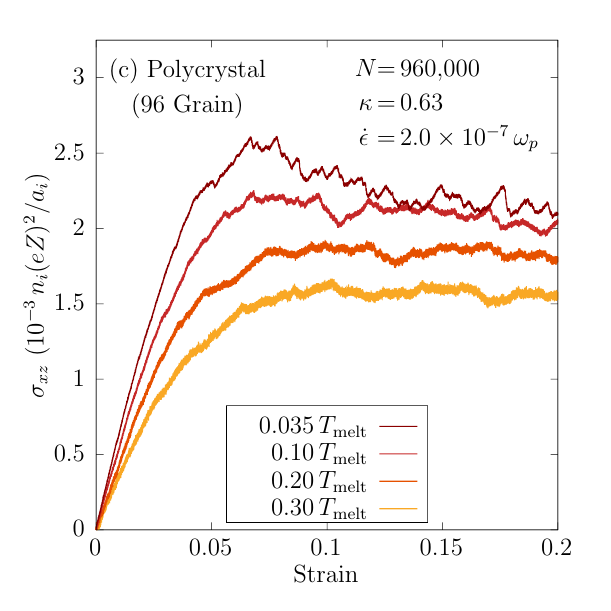}
\caption{\label{fig:stresstrain1} Strain rate dependence for (left) polycrystals and (center) monocrystals. Our fastest rate $5.1\times10^{-4} \ \omega_p$ uses ${v_{\rm box}/v_{\rm el}=0.21}$ from HK09 \cite{horowitz2009breaking}, with $\sigma_{\rm yield} =2\times10^{-3} \, n_i (e Z)^2/a_i$  (dashed). The plastic stress in the polycrystal shows modest temperature dependence (right), but may converge below $0.10 \,T_{\rm melt}$.}
\end{figure*}

We now vary the imposed strain rate on our polycrystal in Fig. \ref{fig:stresstrain1}a. 
When strained sufficiently slowly, these polycrystals robustly show linear elasticity transitioning to plastic flow at $\epsilon \approx 0.05$ and $\dot\epsilon \leq 2\times10^{-5} \, \omega_p$ ${(v_{\rm box}/v_{\rm el} = 8.5\times10^{-3})}$. This critical strain is consistent with recent analytic predictions \cite{BaikoChugunov2018}. 
At converged strain rates, $\sigma_{\rm yield}$ does not show a noticeable dependence on shear rate over the 3 orders of magnitude we simulate. The slope of the stress-strain curve at $\epsilon < 0.05$ is nearly linear so is interpreted as linear elasticity, but the apparent shear modulus (slope) depends weakly on strain rate at fast non-converged strain rates. At small strains the shear modulus in the polycrystal is less than perfect monocrystals by a factor of about 3, between the predictions for polycrystals of Hill and Reuss in ref. \cite{Kobyakov2015}. Simulations extended to $\epsilon = 0.60$ exhibit this flat plastic stress out to very large strains (omitted for length).

Observe that at our highest strain rate the results are poorly converged but qualitatively match the results of Fig. 1 in HK09. From the Orowan equation, stress overshoot is interpreted as a finite time effect when box deformation velocity becomes comparable to the elastic wave speed. Defects such as dislocations and grain boundaries  govern the plastic flow, but if the strain is too fast then they do not have time to nucleate the appropriate defect density to accommodate the imposed flow rate. The amorphization reported in HK09 can perhaps be interpreted as a complete filling of the simulation volume with defects such that it starts to slip like a liquid. This is similar to the interpretation of grain boundaries as thin film liquids in the Fisher model, studied in detail for neutron star crust matter recently \cite{GBD}. 

We now consider a bcc monocrystal. 
Simulations of a perfect monocrystal with $N=128\,000$ in Fig. \ref{fig:stresstrain1}b show perfect linear elasticity up to $\epsilon \approx 0.11$ almost independent of strain rate. For strain rates slower than $\dot\epsilon \leq 2\times10^{-6} \, \omega_p$ we clearly see convergence, with the crystal breaking at $\epsilon \approx 0.11$ as in \citet{Hoffman2012}. For much faster strain rates, the apparent linear elastic regime continues out to larger stresses. Our fastest strain ${(\dot\epsilon = 5.1\times10^{-4} \, \omega_p)}$ shows a turnover at $\epsilon \approx0.15$ 
This is because breaking involves thermal activation to nucleate a defect like a grain boundary to serve as a slip plane. 
The breaking stress in a perfect crystal can be predicted by the Zhurkov model for sufficiently slow deformation, found in MD simulations with similar box sizes to be about $\dot\epsilon  \approx 10^{-5} \, \omega_p$
\cite[see Fig. 1 of ][]{Chugunov2010}. 
For faster strain rates the breaking stress rises rapidly because there is insufficient time for the required thermal activation for a breaking event, explaining the overshoot.

In our slowest simulation, beyond breaking ($\epsilon \gtrsim  0.11$) we observe a grain boundary has formed in the ($\bar{1}01$) plane and slip seems to occur along this surface (simulation animation SM2). The flow stress is again about $2\times10^{-3} \, n_i (eZ)^2/a_i $, 
though the stochasticity is larger because the defect network is much less complicated. This results in larger stress accumulations and larger slips.

We also observe some bi-linear elasticity (inset), where the shear modulus decreases between $0.10 \lesssim \epsilon \lesssim 0.11$ as it approaches the break. This is the onset of stress-activated defect formation, a precursor to an avalanche and catastrophic failure. This bi-linearity may be visible in smaller simulations in Fig. 1 of ref. \cite{Hoffman2012}, but is not discussed by the authors.

We now simulate a range of temperatures typical of neutron star crusts using ${\dot\epsilon =  2\times10^{-7} \, \omega_p}$ in Fig. \ref{fig:stresstrain1}c. These are our largest configurations, with $N=960\,000$ and 96 grains; these grains have comparable size to those in the smaller polycrystal above. The linear elastic shear modulus varies by less than a factor of two over this decade in temperature; $\mu_{\rm eff}$ is almost converged to the zero temperature limit at the temperatures we simulate \cite[\textit{e.g.} Fig. 1 in][]{Strohmayer1991} so this may be evidence of creep at our higher temperatures.
The coldest simulation ($0.035\, T_{\rm melt}$) may slightly overshoot despite our smaller simulation demonstrating convergence at this rate because $v_{\rm box}/v_{\rm el}$ is now a factor of two larger. It is obvious that the stochastic crackling during the plastic regime is merely a finite size effect and with sufficiently large volumes there are always many small breaks happening that can average out to a flat curve. The yield stress has a modest temperature dependence with variation between ${\sigma_y \approx1.5-2\times10^{-3} \, n_i (eZ)^2 / a_i}$, but the stochastic breaking makes it difficult to fit an obvious scaling and requires future work.

\begin{figure}
\includegraphics[trim={12 12 18 16},clip,width=0.98\columnwidth]{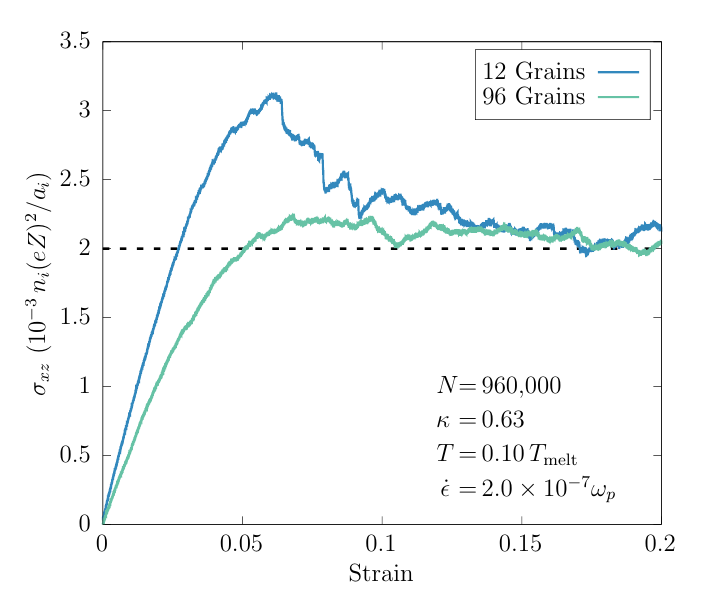}
\includegraphics[width=0.98\columnwidth]{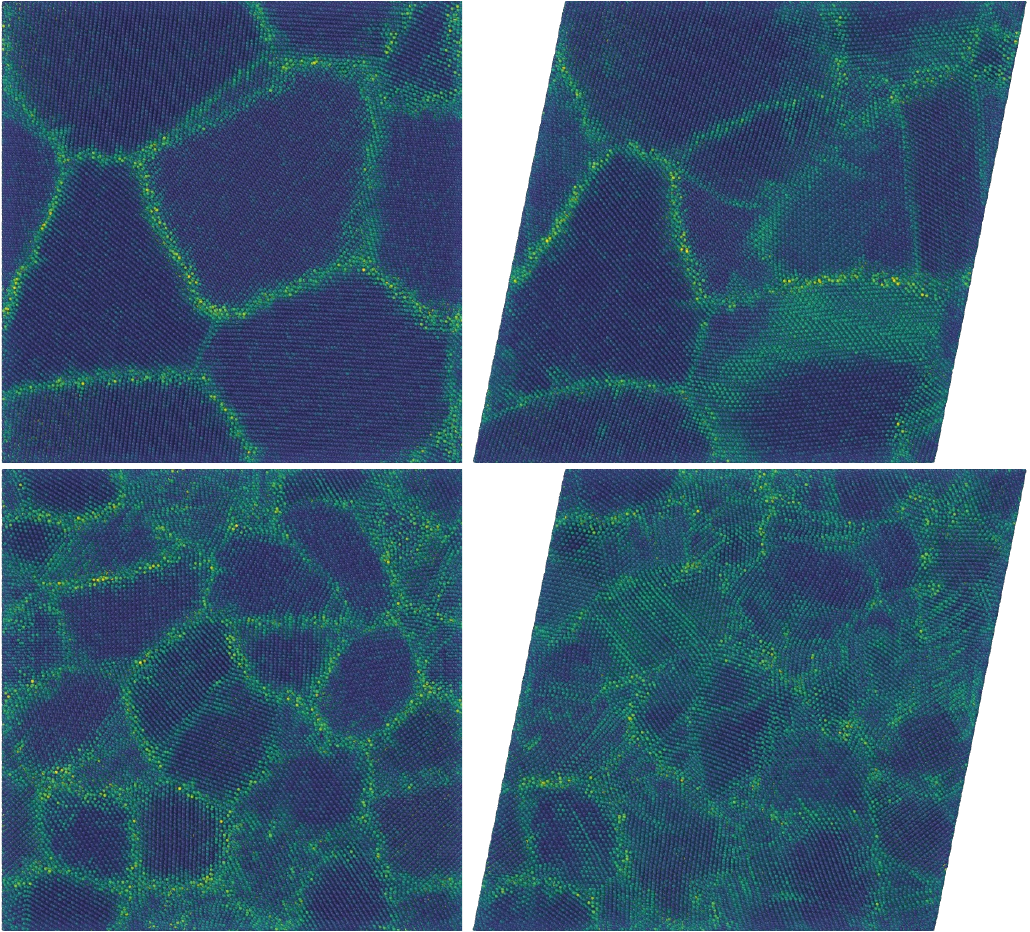}
\caption{\label{fig:million} (Top) Stress-strain curves for configurations with $N=960\,000$ nuclei converge to the same yield stress. Initial (bottom left) and final (bottom right) configurations with (top) 12 grains and (bottom) 96 grains.}
\end{figure}

We study the dependence on initial grain size in Fig. \ref{fig:million} (animation SM3). We perform an additional simulation with $N=960\,000$ at $T=0.10\, T_{\rm melt}$ using ${\dot\epsilon = 2 \times 10^{-7}} \, \omega_p$ ${(v_{\rm box}/v_{\rm el} = 1.7\times10^{-4})}$ but now with 12 grains, effectively varying the initial number density of defects by an order of magnitude relative to Fig. \ref{fig:stresstrain1}c. Both simulations converge to the same post-break plasticity, independent of the initial crystal structure, suggesting this plastic flow is universal. The crystals self-consistently develop a defect network with the appropriate defect density to accommodate the imposed strain rate, losing memory of the initial defect density beyond the critical strain. Though the configuration with 12 grains initially overshoots, this is unsurprising given that the smaller box was only approaching converged plasticity at similar $v_{\rm box} /v_{\rm el}$ above.
Due to the lower volume filling fraction of grain boundaries in the initial configuration, this simulation has a lower defect density and must nucleate many more to accommodate the imposed strain rate.

\textit{Discussion\textemdash} Using large-scale MD simulations we model the deformation and breaking of neutron star crusts in response to large shear strains. By shearing four orders of magnitude slower than past work, we demonstrate converged quasi-static evolution and discover a qualitatively new regime of steady plastic flow beyond the breaking point. The plastic flow shows remarkable independence to the initial crystal structure and asymptotic yield stress $\sigma_{\rm yield} = 1.5 -2\times10^{-3} n_i (eZ)^2/a_i$. This value is easily understood knowing the critical strain $\epsilon\approx 0.05$ and ${\mu_{\rm poly} \approx \mu/3}$, such that $0.11\times0.05/3 \approx 2\times10^{-3}$. 
We argue that slip from defects, especially grain boundaries and dislocations are responsible for the plastic flow, and that the universal post-break plasticity is a result of the crystal self-consistently nucleating an appropriate density of defects to accommodate the imposed strain rate.

Ultimately, if stress accumulates on length scales larger than the typical grain size the crust may flow plastically without sudden transients associated with the catastrophic release of stored elastic energy. However, if gradients in stress fields have smaller characteristic length scales than crustal grains then the crust may yield suddenly like our monocrystal simulations. Magnetically driven plastic flows were proposed to explain magnetar outbursts \cite{Beloborodov_2014,Li_2016,Thompson2017,Kaspi_2017}, while quakes associated with sudden yielding events were considered as a mechanism to trigger X-ray bursts, fast radio bursts, and spin glitches \cite{Thompson_1995,Thompson_1996, yuan2020, Bransgrove_2020}. The MD simulations in this work indicate that the diversity of magnetar activity may be closely related to the distribution of grain sizes in their crusts.

Since initially perfect crystals have a much higher activation threshold for breaking, we consider that neutron star crusts may be prone to cycling, where initially macroscopic crystals store and release large amounts of elastic energy, and then flow plastically \cite{Baiko2024}. If these crystals can then heal and anneal back to large sizes, crust breaking may repeat with implications for starquakes and magnetar bursts. Such crustal healing will depend sensitively on the temperature and creep rate of the crust in the thermally dominated limit of the Orowan equation. Future work will characterize the detailed visco-elastic properties of the crystal and inform macroscopic modeling of the crust and its connection to observed magnetar activity \cite{Li_2016,Thompson2017,Lander_2019,QuBransgrove2025,Burnaz_2025}.  
The fact that these crystals exhibit familiar plastic physics is perhaps unsurprising, but the direct confirmation here now strongly motivates future work to characterize defects such as dislocations in detail; a full deformation mechanism map will enable mesoscale modeling with realistic rheology.

\begin{acknowledgments}
We thank Greg Huber and Karin Dahmen for discussion. 
This work was supported by a grant from the Simons Foundation (MP-SCMPS-00001470) to MC. This research was supported in part by grant NSF PHY-2309135 to the Kavli Institute for Theoretical Physics (KITP). MC and AB thank the KITP for hospitality and MC acknowledges support as a KITP Scholar.  
Financial support for this publication comes from Cottrell Scholar Award \#CS-CSA-2023-139 sponsored by Research Corporation for Science Advancement.  
AB is supported by a PCTS fellowship and a Lyman Spitzer Jr. fellowship.
This research used the DeltaAI advanced computing and data resource, which is supported by the National Science Foundation (award OAC 2320345) and the State of Illinois. DeltaAI is a joint effort of the University of Illinois Urbana-Champaign and its National Center for Supercomputing Applications. CJH acknowledges the support of the US Department of Energy grant DE-FG02-87ER40365 and the National Science Foundation grant PHY 21-16686

\end{acknowledgments}




%

\end{document}